\documentclass[a4paper,10pt]{article}

\usepackage[latin1, utf8]{inputenc}
\usepackage[T1]{fontenc}
\usepackage{gensymb}
\usepackage{authblk}
\usepackage[dvips]{graphicx} %%%%pour l'introduction de figures PS dans le document%%%%%%%

\title{Adaptation of the Bridgman anvil cell to liquid pressure mediums\footnote{Copyright (2007) American Institute of Physics. This article may be downloaded for personal use only. Any other use requires prior permission of the author and the American Institute of Physics. The following article appeared in Rev. Sci. Instrum. \textbf{78}, 123901 (2007) and may be found at http://link.aip.org/link/RSINAK/v78/p123901.}}

\author{A.-S. Rüetschi and D. Jaccard}
\affil{DPMC, University of Geneva, 24 Quai E.-Ansermet, 1211 Genève 4, Switzerland}

\date{}

\begin{document}

\maketitle

%\begin{abstract}

The advantage of Bridgman anvil pressure cells is their wide pressure range and the large number of wires which can be introduced into the pressure chamber. In these pressure cells soft solid pressure mediums like steatite are used. We have succeeded in adapting the Bridgman cell to liquid pressure mediums. With this breakthrough it is now possible to measure in very good hydrostatic pressure conditions up to 7\,GPa, which is about twice the pressure attainable in piston-cylinder cells. The pressure gradient in the cell, estimated from the superconducting transition width of lead, is reduced by a factor of five in the liquid medium with respect to steatite. By the use of non-magnetic materials for the anvils and the clamp and due to the small dimensions of the latter, our device is specially suitable for magneto-transport measurements in dilution fridges. This pressure cell has been developed to measure very fragile and brittle samples like organic conductors. Resistivity measurements of (TMTTF)$_{2}$BF$_{4}$ performed in a solid and liquid pressure medium demonstrate the necessity of hydrostatic pressure conditions for the study of organic conductors at high pressures.

%\end{abstract}

\maketitle

\section{Introduction}

High pressure is an important tool for studying the electronic properties of solids. Pressure can be used as a control parameter in the same way as doping or electric and magnetic fields and can drive systems through phase transitions. A crucial point in all high pressure experiments are the pressure conditions in the cell. They need to be as hydrostatic as possible in order to get more than only qualitative results. The pressure transmitting medium is therefore very important. In piston-cylinder cells, which cover a pressure region up to about 3.5\,GPa, liquids such as oils or alcohols are used as pressure medium. These liquids still conserve good hydrostaticity when they solidify. Helium filled diamond anvil pressure cells (DAC) offer the possibility to achieve much higher pressures than piston-cylinder cells and excellent pressure conditions. However, due to the high compressibility of helium and the small sample size required (length $\sim$ 0.2\,mm), it is very difficult to perform precise transport measurements (e.g. four point resistivity measurements) in such cells \cite{Thomasson_1997, Thomasson_1998, Demuer_2001, Holmes_2004}. Transport measurements in DACs on brittle samples like organic conductors do not seem practicable because mechanically strong wire contacts are needed. Another technique for pressures exceeding 3.5\,GPa are the Bridgman anvil pressure cells. They have proven worthwhile, mainly by offering the possibility to introduce a large number of wires into the pressure chamber. In this type of pressure cells only solid pressure transmitting mediums were used because of the porosity of the pyrophyllite gasket. A widely used material is steatite (or soapstone), a magnesium-silicate talc. This soft mineral is easily machinable and has fairly good pressure properties. Typical pressure gradients in cells with steatite can reach 5-10\% of the pressure in the cell. While this is good enough for a lot of applications, it is not sufficient --- as we show in this article --- for measuring brittle samples. In order to improve the pressure conditions in Bridgman anvil pressure cells, a liquid pressure medium is indispensable. Different techniques have been developed so far. In Russia, the large volume toroid pressure cell has been adapted to liquid mediums by inserting a teflon capsule containing the liquid in the central zone of the cell \cite{Khvostantsev_2004}. Through its large volume this cell is not suited for low temperature measurements. Nakanishi \textit{et al.} \cite{Nakanishi_2002} also use modified Bridgman anvils in which a teflon capsule is placed. The anvils can be put in a clamp sufficiently small for dilution fridges. The drawback of this method is that only few wires can be introduced into the cell. Recently, this device has been adapted for nuclear magnetic resonance (NMR) and nuclear quadrupole resonance (NQR) measurements up to 9\,GPa, where the Cu$_2$O is used as in situ NQR manometer \cite{Fukazawa_2007}. A simpler approach is made by Colombier \textit{et al.} \cite{Colombier_2007}. They use flat anvils which renders complex gasket and teflon forms unnecessary. The cell is formed by two teflon or nylon rings placed inside the gasket. Under pressure the rings seal the cell. We adopt a different approach using no teflon capsule. The advantages of our technique are the large number of wires (up to fourteen, to date), the small dimensions of the clamp and the non-magnetic materials used. The device is specially designed for measurements in dilution fridges. We now have a very versatile technique providing nearly hydrostatic pressure conditions up to 7\, GPa at temperatures as low as 30\, mK and which is compatible with high magnetic fields. In the following we describe in detail our new technique and its preformances. Furthermore we demonstrate its necessity for measurements of very brittle samples by showing the influence of the pressure medium on the resistivity of organic conductors.

\section{Device description}

The design of our pressure cell is based on the classical Bridgman setup \cite{Bridgman_1952}, further developped by Wittig \cite{Wittig_1966}. In this setup, the sample space is formed by a pyrophyllite gasket which is squeezed between the flat parts of two opposed anvils. Pyrophyllite, an aluminium silicate hydroxide, is used for its high friction coefficient with respect to metals and its intermediate internal friction coefficient \cite{Wentorf_1962}. These frictional forces allow the pyrophyllite gasket to withstand the high pressure difference existing across its width, while --- due to the intermediate internal friction --- it is still sufficiently compressible. Since pyrophyllite is a porous material a solid pressure transmitting medium is required. A detailed description of such a pressure cell can be found in \cite{Jaccard_1998}.

The replacement of the solid pressure medium by a liquid in this Bridgman setup involves two major modifications. Firstly, the cell must be sealed. At high pressures this is simply done by the pyrophyllite gasket itself which becomes water-proof as pyrophyllite undergoes a plastic transformation at high pressures. The simplest way to build a pressure cell would therefore be the following: gluing a pyrophyllite gasket on the anvil, fixing it on the outside with epoxy (as is done in the classic setup) and then soaking the whole gasket with the liquid pressure medium. Pressure is generated in such a cell but the pronounced elliptic form of the sample space after pressurisation indicates that the gasket is not very stable. Better results are achieved when the inside of the gasket is covered with an epoxy film. The second change concerns the way samples are contacted. In the classic setup the sample is contained between two discs of the solid pressure medium. The wires are simply placed on the sample and the pressure medium is used to press the wires onto the sample. When a liquid pressure medium is used, these mechanical contacts must be replaced by spot-welded or glued ones.

Anvils made of non-magnetic tungsten carbide (WC) are used. The anvil flat is 3.6\,mm in diameter, the Bridgman angle 6\degree~and the body has a conicity of 4\%. The anvils are press-fitted in copper-beryllium (Berylco 25, CuBe) shrink collars. Through this procedure the anvils are laterally compressed. This stress is opposite to the one generated by the press and therefore the load resistance of the anvils is increased. With this type of anvil, the maximum pressure is limited to about 10\,GPa. Areas of the anvils which could come in contact with either the sample or the wires are covered with an insulating paint.

The pressure cell is directly built on the lower anvil (see fig. 1 and 2). The gasket is made of two pyrophyllite rings with a total height of about 0.220(2)\,mm. Their outer diameter $\phi_{e}$ is adjusted to the anvil flat, the inner diameter $\phi_{i}$ is about 2\,mm. One of these rings with a height of about 0.11\,mm is glued on the anvil with a sodium silicate solution and in addition fixed on the outside with epoxy (as epoxy, we use the fast hardening Araldite). With a razor blade, up to fourteen trenches, which hold the wires, are cut into this ring. The depth of the trenches is about 40\,$\mu$m so that the wires fill the excavated space without protruding, allowing therefore the second ring to be \textit{perfectly} coplanar to the first one. The passage of the wires is done in two steps. First, wires that connect the outside to the cell are wedged in the trenches. These outer wires do not extend into the pressure chamber but end in the gasket. Annealed gold wires with 25\,$\mu$m diameter are used for this purpose. In a second step, annealed gold wires with a diameter of 10\,$\mu$m are connected to the samples. The wires are attached to the samples either by spot-welding or --- if the sample is brittle --- by gluing them with a silver paste. This is done outside the pressure cell. The samples with their wires are then placed in the cell so that all these inner wires fit in the trenches and overlap the outer wires. The electrical contacts between the outer and inner wires establishes in the gasket when the cell is pressurised. To have already a contact at ambient pressure a droplet of silver paste is added just outside the gasket. In order to reduce the mechanical force on the sample contacts during pressurisation, supports are placed in the cell between the samples and the pyrophyllite gasket on which the wires are stuck. The second pyrophyllite ring is placed on top of the first one and fixed on the outside with epoxy. Partially polymerized epoxy is used for this in order to prevent it flowing by capilarity into the wire passages or in between the two pyrophyllite rings. The epoxy helps the formation of the gasket at the beginning of the pressurisation and also insulates the wires from the anvils. The inside of the pyrophyllite gasket is sealed with an epoxy film of about 40\,$\mu$m thickness. Here again, partially polymerized epoxy is used, so that it does not flow over the anvil. The upside of the gasket is covered with the sodium silicate solution.

When using a solid pressure medium, the exact ratio between the volume of the pressure transmitting medium and the volume of the cell is crucial for a good stability of the gasket. A filling factor of about 75\% is optimal for steatite \cite{Jaccard_1998}. An accurate determination of the filling factor is very difficult when a liquid pressure medium is used. We therefore pour an excess of liquid into the cell. The surplus liquid is ejected at the beginning of the pressurisation. The drawback of this method is, that the height of the gasket needs to be reduced in order to ensure a stable gasket. This lowers the maximum attainable pressure. Through a series of test cells with different gasket heights ranging between 0.198\,mm and 0.230\,mm the best efficiency, that is the highest pressure versus applied load coefficient, has been found for a gasket height of 0.220\,mm \cite{footnote1}. It can happen that some liquid is flowing on the upside of the pyrophyllite gasket during the filling of the pressure chamber, but this does not influence the performances of the pressure cell. At the beginning, polyethylene siloxane has been used as pressure medium, but it is suboptimal because it evaporates. Therefore an oil, Daphne Oil 7373 \cite{Idemitsu}, has been selected.

The clamp is a tube made of a non-magnetic titanium-aluminium-vanadium alloy (Ti-6Al-4V, ASTM F136) (see figure 3). Both ends can be closed by screws. For pressurisation the clamp is suspended in a support. The load is applied by the piston of an oil press passing through a hole in the upper screw. The force is blocked by tightening the upper screw. The clamp is designed so that the thread of this screw is stressed when the load is applied, thereby minimising the pressure decrease due to the release of the load. The wires can leave the clamp through four windows which are positioned at the height of the anvils. They are then conducted to the lower end of the clamp in trenches. In order to have isobaric measurements when cooling the device, the load must be kept constant. This is the case when the thermal contractions of the different parts compensate each other. In our setup the anvils (WC) are placed between two supports made of CuBe. The coefficient of thermal contraction of the clamp material (TiAlV) is intermediate to those of WC and CuBe and therefore, by adjusting the dimensions of these parts, pressure variations during cooling can be minimized. A separate copper jacket is used for thermalising the clamp in very low temperature measurements. The windows which are not used for the passage of the wires (two or three, depending on the number of wires) can be filled with copper pieces. The copper jacket presses these pieces against the shrink collars which are in good thermal contact with the anvils. This direct thermal path from the copper jacket to the anvils is essential to achieve low temperatures in the pressure chamber when measuring in the dilution fridge. The electrical leads also help to thermalize the samples. The thermometers are placed in a small copper block which is fixed on the jacket. The small dimensions of the clamp (diameter: 3.0\,cm, height: 11.5\,cm) make this device suitable for different measurement apparatus, e.g. dilution fridges.

\section{Performance}

We determine the pressure in our cells in situ by measuring the superconducting transition temperature of lead. The maximum pressure obtained until now is 7\,GPa for an applied load of 64\,kN. The width of the superconducting transition \cite{footnote2}, which is less than 1\,mK at ambient pressure, reflects the pressure homogeneity in the cell. Figure 4 shows two superconducting transitions of lead, one measured in a liquid pressure medium, the other in a solid one (steatite) at close pressures. The width of the transition is greatly reduced in the liquid pressure medium. Our measurements performed so far in a liquid medium have a mean transition width of 14\,mK. This is about five times smaller than the transitions measured in steatite and corresponds to a pressure variation $\Delta$p of about 0.03\,GPa. The pressure variation only weakly increases with pressure. The small but non negligible pressure inhomogeneity comes from the fact that the pressure transmitting liquid undergoes a vitrous transition at high pressures and low temperatures. At room temperature, Daphne Oil 7373 solidifies at 1.9\,GPa \cite{Itou_2004}. Another convenient way to estimate the pressure homogeneity in a cell is to measure the structural phase transitions of bismuth at ambient temperature (see figure 5) \cite{Pedrazzini}. These transitions are fixed pressure points at 2.54 and 2.70\,GPa for the bismuth I-II and II-III transitions respectively and are accompanied by big jumps in the resistivity. In steatite the pressure width $\Delta$p$_{Bi}$ of these transitions is about 0.1 to 0.2\,GPa. In the liquid pressure medium $\Delta$p$_{Bi}$ is no longer resolvable with our press, which means that $\Delta$p$_{Bi}$\,<\,0.05 GPa.

The pressure variation between 300\,K and 4\,K is about 0.1\,GPa. To measure this, a cell containing a bismuth and a lead sample was built. At room temperature, the pressure was fixed precisely at one of the structural phase transitions of bismuth and then compared with the pressure determined by the superconducting transition temperature of lead \cite{Rolf}. Furthermore, the resistivity curves $\rho(T)$ of lead at different pressures can be scaled onto the curve at ambient pressure. Provided that the phonon spectra does not change under pressure, this shows that the measurements are isobaric.

Samples whose resistivities vary strongly with temperature give useful information about the thermalization, i.e., the ratio of the temperature of the thermometer to the effective temperature of the samples. Above the superconducting transition the lead manometer serves as an in situ thermometer. The thermalization is reasonable down to 10 K. Around 4.2\,K the lag between the thermometer and sample response is only a few seconds. Below 1\,K the thermalization is slow. The lowest temperature reached with our device is 30\,mK in continuous dilution.

All cells are carefully depressurized. The cells which have been driven to the maximum load of 64\,kN exploded at an applied force of about 20\,kN due to gasket instabilites. But a full pressure cycle is possible up to at least 30\,kN ($\simeq$ 2.5\,GPa). The hysteresis between loading and unloading is 10-20\,kN and is due to the irreversible plastic deformation of the gasket. The hysteresis ends through a sudden pressure decrease at which point the cells driven to maximum load explode. After depressurizing the cell without explosion the samples are found in their initial shape and size. Even the very brittle organic conductors show no deformation. This is in complete contrast to steatite cells, where the organic conductors are found to be seriously damaged after the opening of the cell. Under a pressure of 7\,GPa the inner diameter $\phi_{i}$ decreases to about 1.8\,mm and the gasket height h to about 160\,$\mu$m. We therefore have a large working space available. This and the large number of wires make it possible to measure the resistivity of up to four samples with a four probe method or to design various complex setups.

\section{Application to organic conductors}

The driving motivation to develop this pressure cell has been the desire to study the Fabre and Bechgaard salts. These organic conductors are isostructural compounds. Flat organic molecules (TMTTF for Fabre salts, TMTSF for Bechgaard salts; TM short for both) are piled up in the a-direction. These stacks are aligned with each other along the b-direction and separated by anions in the c-direction. Two TM molecules transfer one electron to each anion which then have a closed shell and do not contribute to the transport. Through this crystal structure the TM salts have very anisotropic properties and can be considered as weakly coupled conducting chains. The resistivity along the a-axis, $\rho_a$, for instance, is two orders of magnitude lower than $\rho_b$ and more than three orders of magnitude lower than $\rho_c$. The Fabre and Bechgaard salts present a rich temperature versus pressure phase diagram. This phase diagram has been found to be generic for all TM salts \cite{Wilhelm_2001}. Instead of applying external pressure, the chemical pressure can be varied by changing the anions. The different compounds can therefore be placed on the pressure axis corresponding to their ambient pressure properties. At low pressures and high temperatures the TM salts are quasi-one-dimensional conductors. By lowering the temperature and increasing the pressure the salts cross over to a more two- and three-dimensional conductor. TM salts have different ground states depending on composition and pressure (spin-Peierls, antiferromagnetic, spin-density-wave, superconducting and metallic). By applying pressure, the salts positioned at the low pressure side are driven through a cascade of such ground states. For a review on TM salts see e.g. \cite{Toyota_2007}.

Measuring organic conductors under pressure is not easy because they are extremely brittle and therefore easily crack if the pressure is not uniform. In this section we show the influence of the pressure medium on the a-axis resistivity $\rho_{a}$ of (TMTTF)$_{2}$BF$_{4}$ single crystals. Typical sample sizes are about a\,$\times$\,b$^\prime$\,$\times$\,c$^\star$ = 0.6$\times$0.06$\times$0.05\,mm \cite{footnote3}. The resistivity is measured with a standard four probe method. Four gold rings have been evaporated onto the samples by P. Auban-Senzier (LPS, Université Paris-Sud, Orsay, France) and 10\,$\mu$m gold wires were glued on them with Dupont conductor paste 4929N. Despite the anisotropic transport properties the current is injected homogeneously in the whole sample since the resistivity is unchanged when the current and voltage wires are exchanged. In the calculations of the resistivities the change of the geometrical factors due to the sample compressibility and the contraction on cooling has been neglected ($\rho = (s/l)R$, where s is the sample section and l the distance between the voltage probes, $s/l$ is the geometrical factor). Figure 6 shows the resistivity $\rho_{a}$ of (TMTTF)$_{2}$BF$_{4}$ at room temperature between ambient pressure and 6.5\,GPa measured in a solid (steatite) and liquid (Daphne Oil) pressure medium. Since in the cell with the solid medium the electrical contacts are only established through pressure, there is no measurement of the resistivity at ambient pressure. Resistivities of organic conductors at ambient conditions vary considerably from one sample to another. Nevertheless the resistance at ambient pressure of the sample measured in steatite is probably lower than at low pressures. Small rearrangements of the steatite, which are unavoidable at the beginning of the pressurisation, damage the sample. This changes the geometrical factor in an unknown manner. The cracks in the sample also induce a grain boundary contribution to the resistivity responsible for the higher resistivity in the sample measured in steatite. After the initial damaging of the sample in the steatite cell, no more cracks appear as shows the smooth decrease of the resistivity. The resistivities of both samples decrease with increasing pressure, the decrease in the sample measured in liquid being twice as large as the one measured in steatite. Even more obvious is the different behavior of the two $\rho_{a}(T)$-curves (figure 7). In steatite, the resistivity decreases slightly at high temperatures before going through a shallow minimum at 160\,K and increasing by more than one order of magnitude as the temperature is lowered. Below 0.2\,K the resistivity shows a tendency to saturate. The sample measured in the liquid pressure medium however remains metallic down to 8\,K with a residual resistivity ratio of about 6. Below 8\,K, a small localization sets in, the resistivity increases by about 4\% and saturates below 0.12\,K. The cracks in the sample measured in steatite clearly inhibit the developpement of the metallic phase. Whether the small localization in the sample measured in the liquid pressure medium is inherent or also due to some minor deformations remains open. This comparison shows that measurements of the brittle organic conductors in a solid pressure medium can give at best a qualitative idea of the sample properties. To obtain reliable values, measurements in a liquid pressure medium are indispensable.

\section{Acknowledgments}

We would like to thank E. Colombier and D. Braithwaite (CEA, Grenoble, France) for fruitful discussions and P. Auban-Senzier (LPS, Université Paris-Sud, Orsay, France) for providing us the prepared organic samples.

\newpage

\begin{figure}[htbp]
 %h=here, t=top, b=bottom, p=separate figure page
  \begin{center}
  \includegraphics{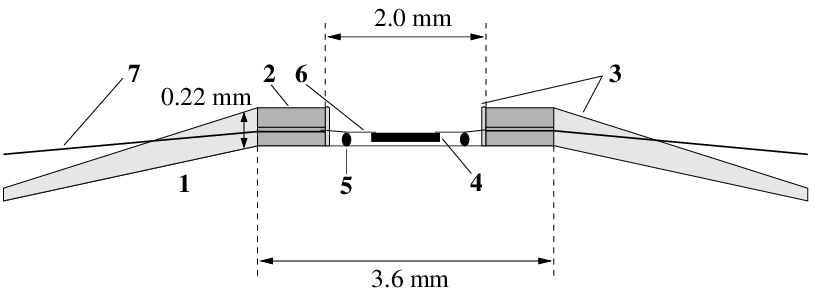}
  \caption { Sectional drawing of the pressure cell. The scale of the vertical axis is doubled with respect to the horizontal axis. 1: anvil, 2: gasket formed by lower and upper pyrophyllite ring, 3: epoxy, 4: sample, 5: support for wire, 6: 10\,$\mu$m gold wire, 7: 25\,$\mu$m gold wire.}
  \end{center}
\end{figure}

\begin{figure}[htbp]
 %h=here, t=top, b=bottom, p=separate figure page
  \begin{center}
  \includegraphics{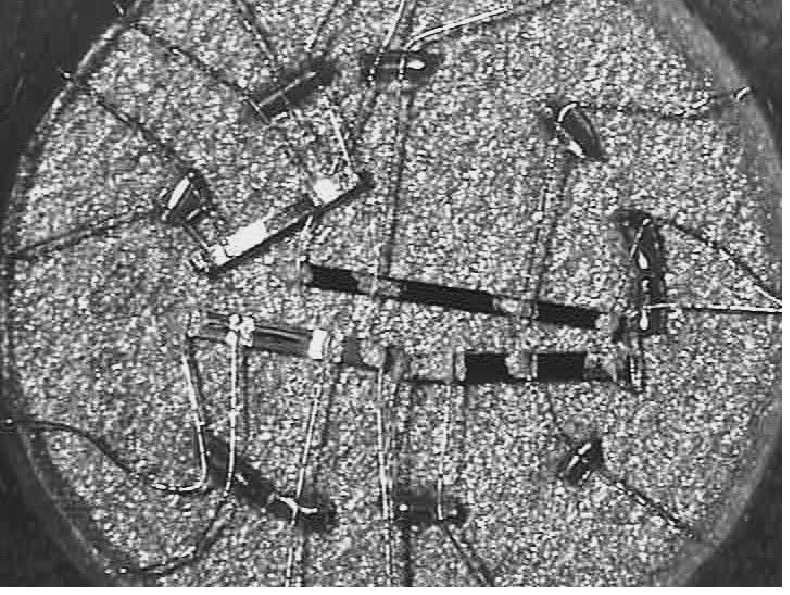}
  \caption { Top view of the pressure cell containing three organic conductors and a lead manometer (in the top left corner). }
  \end{center}
\end{figure}

\begin{figure}[htbp]
 %h=here, t=top, b=bottom, p=separate figure page
  \begin{center}
  \includegraphics{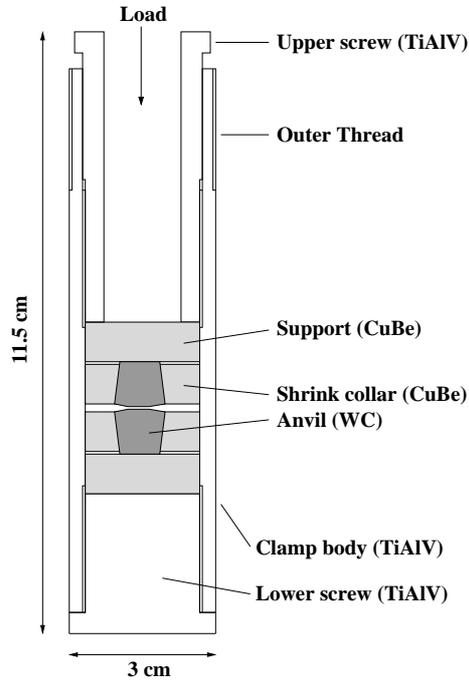}
  \caption { Sectional drawing of the clamp. The outer thread is used to suspend the clamp in a support for pressurisation. }
  \end{center}
\end{figure}

\begin{figure}[htbp]
 %h=here, t=top, b=bottom, p=separate figure page
  \begin{center}
  \includegraphics{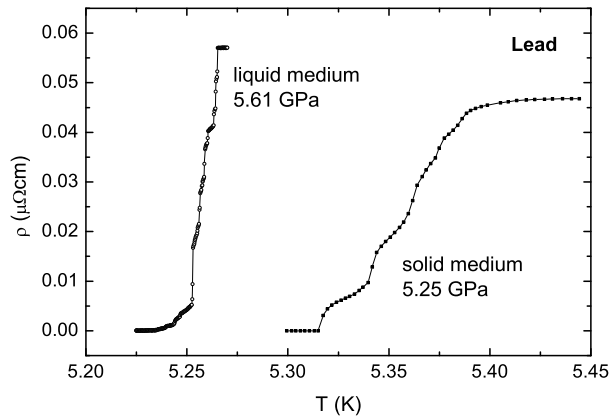}
  \caption { Superconducting transition of lead measured in a solid (steatite) and liquid (Daphne oil) pressure medium. The midpoint value defines the pressure, the width of the transition the pressure gradient in the cell. }
  \end{center}
\end{figure}

\begin{figure}[htbp]
 %h=here, t=top, b=bottom, p=separate figure page
  \begin{center}
  \includegraphics{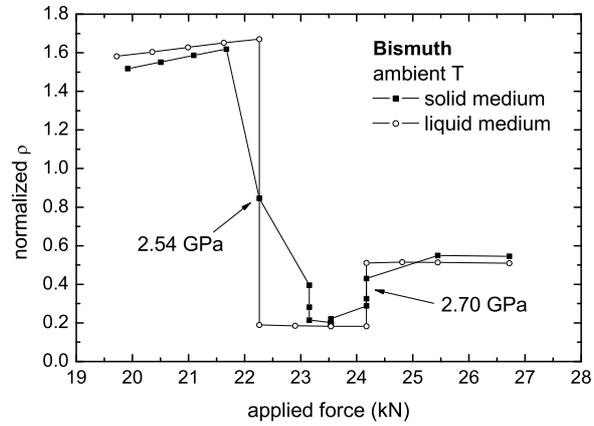}
  \caption { Pressure dependence of the resistivity of bismuth at ambient temperature measured in steatite (closed symbols) and Daphne oil (open symbols). The resistivity is normalized at ambient pressure. The pressure is given as the force developed by the press. The applied force for the steatite cell is normalized to the force applied for the liquid cell. }
  \end{center}
\end{figure}

\begin{figure}[htbp]
 %h=here, t=top, b=bottom, p=separate figure page
  \begin{center}
  \includegraphics{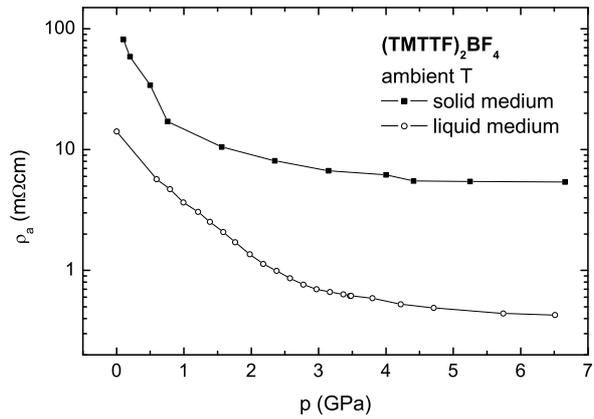}
  \caption { Pressure dependence of the a-axis resistivity of the organic compound (TMTTF)$_{2}$BF$_{4}$ measured in steatite (closed symbols) and Daphne oil (open symbols). }
  \end{center}
\end{figure}

\begin{figure}[htbp]
 %h=here, t=top, b=bottom, p=separate figure page
  \begin{center}
  \includegraphics{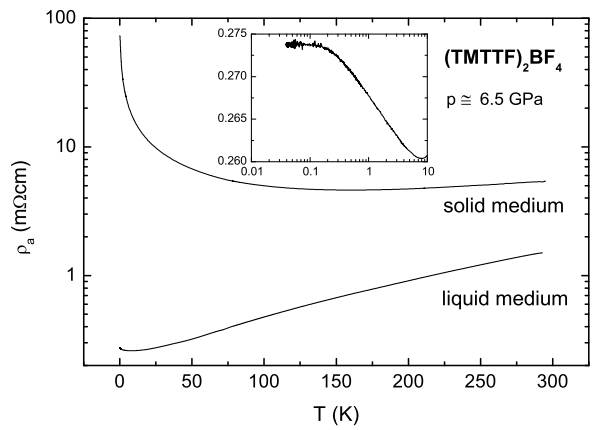}
  \caption { Temperature dependence of the a-axis resistivity of the organic compound (TMTTF)$_{2}$BF$_{4}$ measured in steatite and Daphne oil. The pressure is 6.6\,GPa in the solid medium and 6.51\,GPa in the liquid medium. }
  \end{center}
\end{figure}


\begin{thebibliography}{10}
\bibitem{footnote1}
A high efficiency was obtained with h = 0.25\,mm and partial filling, but the filling control was unsufficiently reliable.
\bibitem{footnote3}
TM-salts are triclinic. b$^\prime$ and c$^\star$ are the projections of the b-and c-axis onto a rectangular parallelepiped.
\end{thebibliography}
\end{document}